\documentclass[9pt]{article}
\usepackage{spconf,amsmath,graphicx}
\usepackage{subcaption}
\usepackage{hyperref}
\usepackage{epsfig}
\usepackage{psfrag}

\makeatletter
\def\blfootnote{\xdef\@thefnmark{}\@footnotetext}
\makeatother

\title{Ultra Low Complexity Deep Learning Based Noise Suppression}
\name{Shrishti Saha Shetu, Soumitro Chakrabarty, Oliver Thiergart, Edwin Mabande}
\vspace{-.1cm}
\address{Fraunhofer IIS, Am Wolfsmantel 33, 91058 Erlangen, Germany \\
\vspace{-.1cm}
\small \textit{\{shrishti.saha.shetu, soumitro.chakrabarty, oliver.thiergart, edwin.mabande\}@iis.fraunhofer.de}}

\begin{document}
%\ninept
%
\maketitle
\begin{abstract}

%\psfrag{[Noisy]}{Noisy}%
%\psfrag{[DeepFilterNet]}{DeepFilterNet}%
%\psfrag{[DeepFilterNet2]}{DeepFilterNet2}%
%\psfrag{[ULCNet$_{MS}$]}{\textbf{ULCNet$_{\text{MS}}$}}%
\psfrag{ULCNet$_{Freq}$}{\textbf{ULCNet$_{\text{Freq}}$}}
This paper introduces an innovative method for reducing the computational complexity of deep neural networks in real-time speech enhancement on resource-constrained devices. The proposed approach utilizes a two-stage processing framework, employing channelwise feature reorientation to reduce the computational load of convolutional operations. By combining this with a modified power law compression technique for enhanced perceptual quality, this approach achieves noise suppression performance comparable to state-of-the-art methods with significantly less computational requirements. Notably, our algorithm exhibits 3 to 4 times less computational complexity and memory usage than prior state-of-the-art approaches.  
  %The increasing complexity of deep neural networks (DNN) for speech enhancement poses challenges for real-time applications on resource-constrained devices. This paper presents a novel approach to address these challenges by proposing a two-stage processing framework, leveraging channel-wise input feature reorientation, to reduce the computational complexity of convolutional operations.  The results demonstrate, that incorporating the proposed new power law compression for improving the subjective perceptual quality along with channel-wise input feature reorientation, the proposed method achieves comparable or even superior noise suppression performance compared to state-of-the-art (SOTA) approaches. Our proposed model has a computational complexity of just 0.098 MACS(G) with only 688K parameters and achieves 12.7$\%$ real time factor (RTF) on a single core of A53 processor.
\end{abstract}


\section{Introduction}
\label{sec:intro}
\vspace{-0.1cm}
\setlength{\belowdisplayskip}{2pt} \setlength{\belowdisplayshortskip}{2pt}
\setlength{\abovedisplayskip}{2pt} \setlength{\abovedisplayshortskip}{2pt}

Speech enhancement (SE) aims to suppress interfering noise while preserving intelligibility and improving the perceived quality of the speech signal. In recent years, there has been a notable rise in the prominence of deep neural network (DNN) based techniques, which have demonstrated a substantial performance improvement over conventional signal processing-based methods \cite{Valin2020APA,Pascual2017SEGANSE,6409417,1163209,1164550}.
State-of-the-art (SOTA) approaches often employ encoder-decoder DNN architectures, reminiscent of U-Net \cite{Choi2021RealTimeDA,hu2020dccrn,choi2020phase,braun2021towards}, operating in the time-frequency (TF) domain. Most of these approaches demand significant computational resources due to the high number of frequency bins in the TF domain.

To overcome this challenge, various techniques have been proposed including channelwise frequency splitting \cite{zhao2022frcrn}, multiband processing \cite{yu2022dmf,10023174}, and using analysis filterbanks inspired by human auditory perception \cite{anderson1984speech}. PercepNet \cite{Valin2020APA} and DeepFilternet2 \cite{schroter2022deepfilternet2},  address this challenge by employing a triangular equivalent rectangular bandwidth (ERB) filter bank. However, this approach only allows for estimating real-valued gains for the ERB bands. To further enhance the phase component of speech, a comb filter, and a deep filtering method \cite{schroter2020clcnet,mack2019deep} were additionally applied. 

%In this approach, the frequency bins present in the magnitude spectrogram undergo logarithmic compression, resulting in 32 distinct ERB bands. In \cite{schroter2022deepfilternet,schroter2022deepfilternet2}, authors proposed a two-stage framework, wherein the first stage similar to \cite{Valin2020APA}, real-valued ERB gains are estimated using an encoder-decoder architecture. In the second stage,

However, despite these recent advancements in reducing computational complexity, deploying these algorithms on resource-constrained embedded devices remains challenging because this typically comes at the cost of compromising on the noise suppression performance. Therefore, in this study, inspired by \cite{schroter2022deepfilternet2, schroter2022deepfilternet}, we propose a two-stage processing framework to address this issue. In the first stage, we employ a highly efficient convolutional recurrent network (CRN) architecture \cite{tan2018convolutional} to estimate an intermediate real magnitude mask. In the second stage, with the help of an even smaller and less complex convolutional neural network (CNN), we enhance the phase components of speech by combining the noisy phase information with the intermediate real magnitude mask. To reduce the computational complexity of the convolution operations in our CRN architecture, we employ the channelwise feature reorientation method described in \cite{liu2020channel}. Unlike common literature approaches \cite{Choi2021RealTimeDA,hu2020dccrn}, we opt not to include a decoder for magnitude mask estimation from bottleneck features. Further, to ensure the robustness of the input features and training targets, we also use a modified power law compression method. 

Our study establishes that our proposed method achieves noise suppression performance comparable to, or even surpassing most SOTA approaches. Our proposed DNN model has only 688K parameters and achieves 0.127 real time factor for a 16 kHz sampling rate on a single core of a Cortex-A53 1.43 GHz processor with a complexity of 0.098 GMACS. This renders our approach highly suitable for deployment on resource-constrained embedded devices.

%With the reduced model size with only 688K parameters and 0.098 GMACS computational complexity, our proposed method achieves 12.7$\%$ RTF on a single core of a A53 processor. 
\vspace{-0.1cm}
\section{Methods}
\vspace{-0.1cm}
\label{sec:methods}
\setlength{\belowdisplayskip}{2pt} \setlength{\belowdisplayshortskip}{2pt}
\setlength{\abovedisplayskip}{2pt} \setlength{\abovedisplayshortskip}{2pt}

%\begin{align}
	%X(n,k) = S(n,k) + N(n,k)
	%\label{eq:NoisySpeech}
%\end{align}
\vspace{-0.2cm}
\subsection{Input Preprocessing}
\vspace{-0.1cm}
\label{sec:IP}
\begin{figure}[t]

\centering
\includegraphics[width=0.47\textwidth]{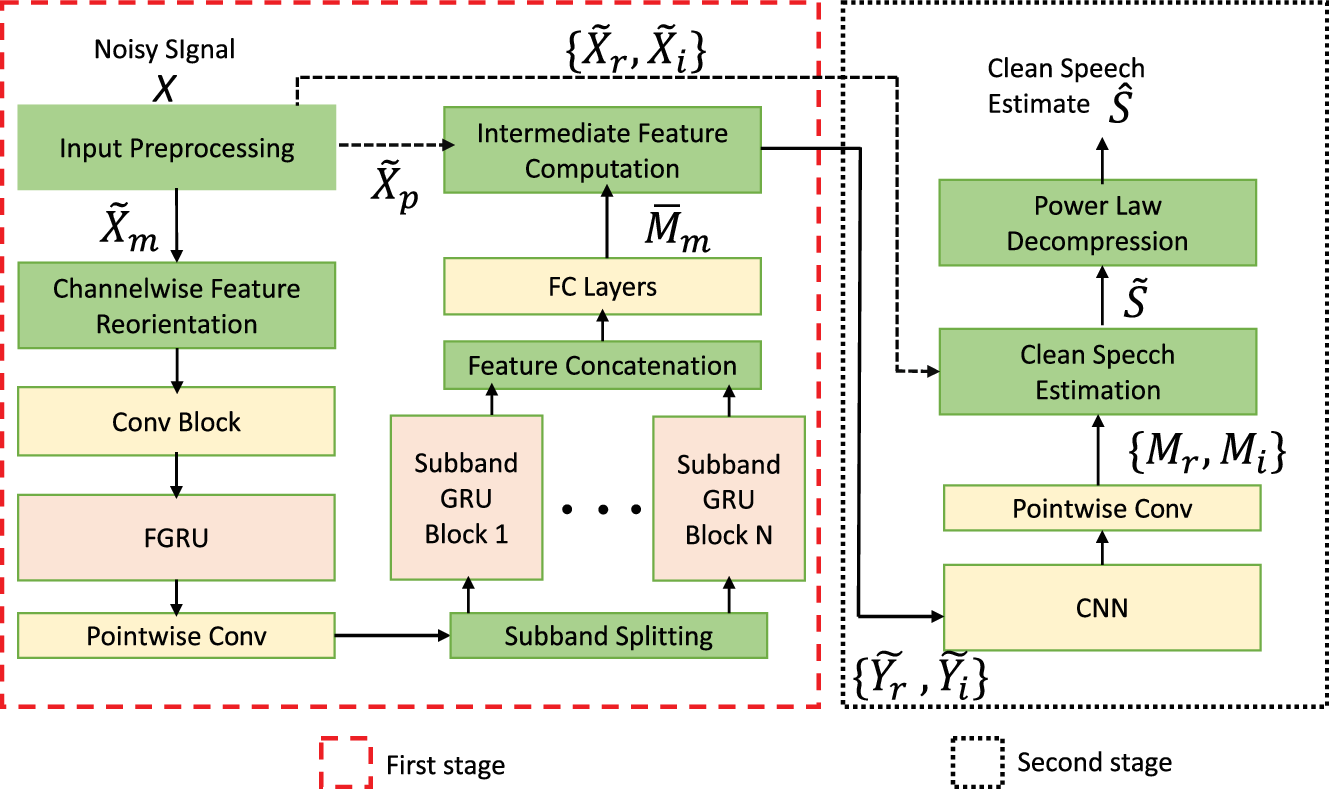}
\caption{\small Proposed Ultra Low Complexity DNN Model}
\label{fig:ULC}
\vspace{-.5cm}
\end{figure}

We assume an additive signal model in the STFT domain where, \(X(n,k)\), \(S(n,k)\), and \(N(n,k)\) denote the noisy signal, the clean speech signal, and the noise components, and $n,k$ denote the time-frame and frequency-bin indices, respectively. 

In our work, the input features provided to the DNN are the magnitude and phase features computed from the power law compressed real and imaginary parts of the noisy signal's STFT. The power law compression method is detailed below. The complex-valued noisy signal $X$ can be expressed in complex rectangular and polar coordinates as:
\begin{align}
X(n,k) = X_{r}(n,k) + jX_{i}(n,k)=\lvert X(n,k)\lvert{e^{j\theta_X}}
\label{eq:RP}
\end{align}
where $X_{r}(n,k), X_{i}(n,k),\lvert X(n,k)\lvert,\theta_X $ denote the real, imaginary, magnitude, and phase part of the noisy signal, respectively. Please note that $(n,k)$ is omitted for brevity in the rest of this paper. 

Normally, in literature, power law compression is applied only to the magnitude part $\lvert X \rvert$ with a factor $\alpha \in [0,1]$ \cite{li2021importance}.   However, in our work, we intend to directly estimate the real ($S_r$) and imaginary ($S_i$) components of the clean speech signal $S$ through implicit complex mask estimation from the DNN. Therefore, we apply the power law compression separately to the real and imaginary parts of the noisy signal $X$ with a power law factor of $\alpha$, as follows:
\begin{align}
&\tilde{X}_{r} = \text{sign}(X_{r}) |X_{r}|^{\alpha} ;   &\tilde{X}_{i} = \text{sign}(X_{i}) |X_{i}|^{\alpha} 
\label{eq:powerlaw}
\end{align}
where the $\text{sign}(X_{r/i})$ denotes that the original sign of $X_{r/i}$ is retained for $\tilde{X}_{r/i}$. Subsequently, the power law compressed noisy magnitude spectrogram and phase component of the noisy signal are obtained as follows:
\begin{align}
&\tilde{X}_{m} = \sqrt{\tilde{X}_{r}^2 + \tilde{X}_{i}^2} ; &\tilde{X}_{p} = \arctan{\frac{\tilde{X}_{i}}{\tilde{X}_{r}}} 
\label{eq:Input Feature}
\end{align}
This $\tilde{X}_{m}$ and $\tilde{X}_{p}$ are then used as the input features to the DNN model.
\vspace{-.2cm}
\subsection{Two Stage Processing and Clean Speech Estimation}
\vspace{-.1cm}
\label{sec:TSP}
\begin{table*}%[t]
\centering
\caption{\small Objective and DNSMOS \cite{reddy2021dnsmos} results on Voicebank+Demand \cite{valentini2016investigating} and DNS challenge 2020 synthetic non-reverb test set. Unreported values of related work are indicated as “-”. RTF, MACS, Params, and objective metrics for Voicebank+Demand are reported from \cite{schroter2022deepfilternet2}.}
\vspace{-.2cm}
\footnotesize
\begin{tabular}{|l|c|c|c|c|c|c|c|c|c|}
\hline
\multicolumn{4}{|c|}{} & \multicolumn{2}{c|}{Voicebank+Demand } & \multicolumn{4}{c|}{DNS Challenge}\\
\cline{5-10}

\textbf{Processing Method}  & \textbf{Params[M]} & \textbf{GMACS} & \textbf{ RTF}& \textbf{PESQ}& \textbf{SI-SDR}&\textbf{PESQ}&\textbf{SI-SDR}&\textbf{SIGMOS}&\textbf{BAKMOS}\\
\hline
Noisy &  - & -& -  & 1.97 & 8.41& 1.58& 9.06& 3.39& 2.61\\
NSNet2 \cite{braun2021towards} &  6.17 & 0.43& 0.02  & 2.47 & -& -& -& -& -\\
% % \hline
PercepNet \cite{Valin2020APA}& 8.00& 0.80&-  & 2.73 & -& -& -& -& -\\
% \hline
%DCCRN  & 3.70& 14.36& 2.19  & 2.54 & -& -& -\\
% \hline
%DCCRN+ &3.30&-& -  & 2.84& -& -& -\\
% \hline
FullSubNet+ \cite{chen2022fullsubnet+} &  8.67 & 30.06& 0.55 & 2.88 & \textbf{18.64}& \textbf{2.99}& \textbf{18.21}& 3.50& 3.91\\
% \hline
DeepFilterNet \cite{schroter2022deepfilternet} &  1.78 &0.35& 0.11  & 2.81 & 16.63& 2.50& 16.17& 3.49& 4.03\\
% \hline
DeepFilterNet2 \cite{schroter2022deepfilternet2} &  2.31&0.36& 0.04 &  \textbf{3.08} & 15.71& 2.65& 16.60& \textbf{3.51}& \textbf{4.12}\\
% \hline
\textbf{Proposed ULCNet$_{\text{MS}}$}&  \textbf{0.688} & \textbf{0.098}&\textbf{0.02}   & 2.87 & 16.89& 2.64& 16.34& 3.46& 4.06\\
\textbf{Proposed ULCNet$_{\text{Freq}}$}&  \textbf{0.688} & \textbf{0.098}& \textbf{0.02} & 2.54 & 17.20& 2.24& 16.67& 3.38& 4.09\\
\hline
\end{tabular}
\label{tab:Objectiveresult}
\vspace{-.3cm}
\end{table*}

In this study, we present a two-stage processing framework for estimating clean speech signal from noisy input. The first stage operates in the magnitude domain, taking $\tilde{X}{_m}$ as input (as explained in section \ref{sec:IP}). As the magnitude spectrogram is considered to be more significant for speech enhancement \cite{tan2018convolutional,wang2014training}, we employ a more computationally intensive CRN-based architecture for this stage, as depicted in Fig. \ref{fig:ULC}. This yields an intermediate real magnitude mask $\bar{M}{_m} \in [0,1]$.

As an input to the second stage, we combine this intermediate real magnitude mask $\bar{M}_{m}$ with the noisy phase $\tilde{X}_{p}$ for intermediate feature computation, as follows:
\begin{align}
&\tilde{Y}_{r} = \bar{M}_{m} *  \cos{\tilde{X}_{p}} ; &\tilde{Y}_{i} = \bar{M}_{m} *  \sin{\tilde{X}_{p}}
\label{eq:IntermediateFeature}
\end{align}
Here, $\tilde{Y}_{r}$ and $\tilde{Y}_{i}$ denote the intermediate real and imaginary feature representation. We also experimented with intermediate feature computation by combining the mask multiplied noisy magnitude with the noisy phase. However, in our experiments, intermediate feature representation as shown in \eqref{eq:IntermediateFeature}, was found to be more beneficial for the second stage. 

%Intermediate feature computation in the signal domain can be shown as follows,
%%%\begin{align}&\bar{Y}_{r} =\tilde{X}_{m}* \bar{M}_{m} *  \cos{\tilde{X}_{p}} ; &\bar{Y}_{i} =\tilde{X}_{m}* \bar{M}_{m} *  \sin{\tilde{X}_{p}}\label{eq:powerlaw}\end{align}
Subsequently, $\tilde{Y}_{r}$ and $\tilde{Y}_{i}$ are concatenated along the channel dimension (as shown in Fig. \ref{fig:ULC}) and processed through a less computationally complex CNN architecture (compared to the CRN architecture in the first stage) to estimate a complex mask $M$.

To estimate the clean speech signal $S$, we apply the complex ratio mask (CRM)-based multiplication method between the noisy signal $\tilde{X}$ and the estimated complex mask $M$, following the approach in \cite{hu2020dccrn}. However, as we applied power law compression to $\tilde{X}_{r}$ and $\tilde{X}_{i}$, as shown in \eqref{eq:powerlaw}, power law decompression is required for the estimated $\tilde{S}_{r}$ and $\tilde{S}_{i}$ to obtain the final clean speech estimate $\hat{S}$, as follows:

\begin{align*}
&\hat{S}_{r/i} = \text{sign}(\tilde S_{r/i})|\tilde{S}_{r/i}|^{1/\alpha};&\hat{S} = \hat{S}_{r} + j\hat{S}_{i}
\end{align*}

%\begin{align*}&F =  \frac{\tanh{M} }{\sqrt{M_{r}^2 + M_{i}^2} }\\&\tilde{S}_{r} = ( \tilde{X}_{r}*M_{r} - \tilde{X}_{i}*M_{i} ) * F \\&\tilde{S}_{i} = ( \tilde{X}_{r}*M_{i} + \tilde{X}_{i}*M_{r}) * F \end{align*}
\vspace{-.2cm}
\subsection{Model Design}
\label{sec:DNNModel}
\vspace{-.1cm}

As detailed in section \ref{sec:TSP}, our work employs a two-stage processing framework. The initial stage utilizes a CRN-based architecture, while the subsequent stage integrates a CNN architecture (see Fig. \ref{fig:ULC}). In the convolutional part of the CRN architecture, we design a Conv block consisting of four 2D depthwise separable convolutional layers \cite{howard2017mobilenets}. The aim of this Conv block is to downsample the input features along the frequency axis and do efficient feature extraction \cite{Choi2021RealTimeDA}. 

Despite employing separable convolutions in the Conv block, we found that convolution operations remain the most computationally expensive part of the CRN architecture. To address this, we utilize the channelwise feature reorientation method introduced in \cite{liu2020channel}. This approach has been proven beneficial for improving the voice separation performance 
on high-resolution music and reducing the computational cost. In our experiments, we found that using channelwise feature reorientation alongside separable convolutions in the Conv block reduces computational complexity by over 5 times compared to the conventional convolution-based Conv block with a similar setup. 

 Following the Conv block, a frequency-axis bidirectional Gated Recurrent Unit (FGRU) \cite{Choi2021RealTimeDA} is used to increase the receptive field and facilitate information sharing along the frequency axis, which is succeeded by a point-wise convolution layer. This bottleneck feature is then divided into subbands and processed by subband temporal GRU \cite{braun2021towards,9747888} blocks. We derive the intermediate magnitude mask $\bar{M}_{m} \in [0,1]$ using only two fully connected (FC) layers, as depicted in Fig. \ref{fig:ULC}. This approach replaces the need for a complete decoder architecture, typically used to upsample bottleneck feature representations after the subband GRU blocks. 

In the second stage, the intermediate feature representation undergoes processing by a CNN architecture featuring two convolutional layers to finally derive the complex mask $M$. Notably, this CNN-based second stage accounts for merely $0.5\%$ of the total parameters of the DNN model shown in Fig. \ref{fig:ULC}.

%We use an overlapping rectangular uniform window to extract the subband features, which are then stacked in channel dimensions.  

%an analysis filter bank is used to obtain the subband features before stacking them in channel dimension. However, in our, we don't use any analysis filter bank. We use an overlapping rectangular uniform window with a frequency resolution of 1.5 kHz (48 STFT bins) with an overlapping factor 0f 0.33 to get each subband. In  total, we get 8 subbands, which are then stacked in channel dimensions [shown in]. 

%In the first stage of the network, we use four 2-D depthwise separable convolutional layers with a kernel size of (1,3). Apart from the first convolutional layer, we use always a max pooling layer with a kernel of (1,2) to further downsample the input feature. The output channel of each of these convolutional layers is {32,64,96,128}.   
%with 64 units and $(1\times1)$ point-wise convolution layer with 64 output filters is used to restrict the bottleneck feature dimension to 384D.

%we use three traditional 2-D convolutional layers with 32 filters and a kernel size of (1,3) to further jointly optimize the magnitude and phase component of the CRM mask. Finally, a $(1\times1)$ point-wise convolution layer with two output channels is used to obtain the desired dimension.
\vspace{-0.3cm}
\section{Experiments and Results}
\label{sec:pagestyle}

\begin{figure*}[t]
\small
    \begin{subfigure}{0.5\textwidth}
        \epsfig{file=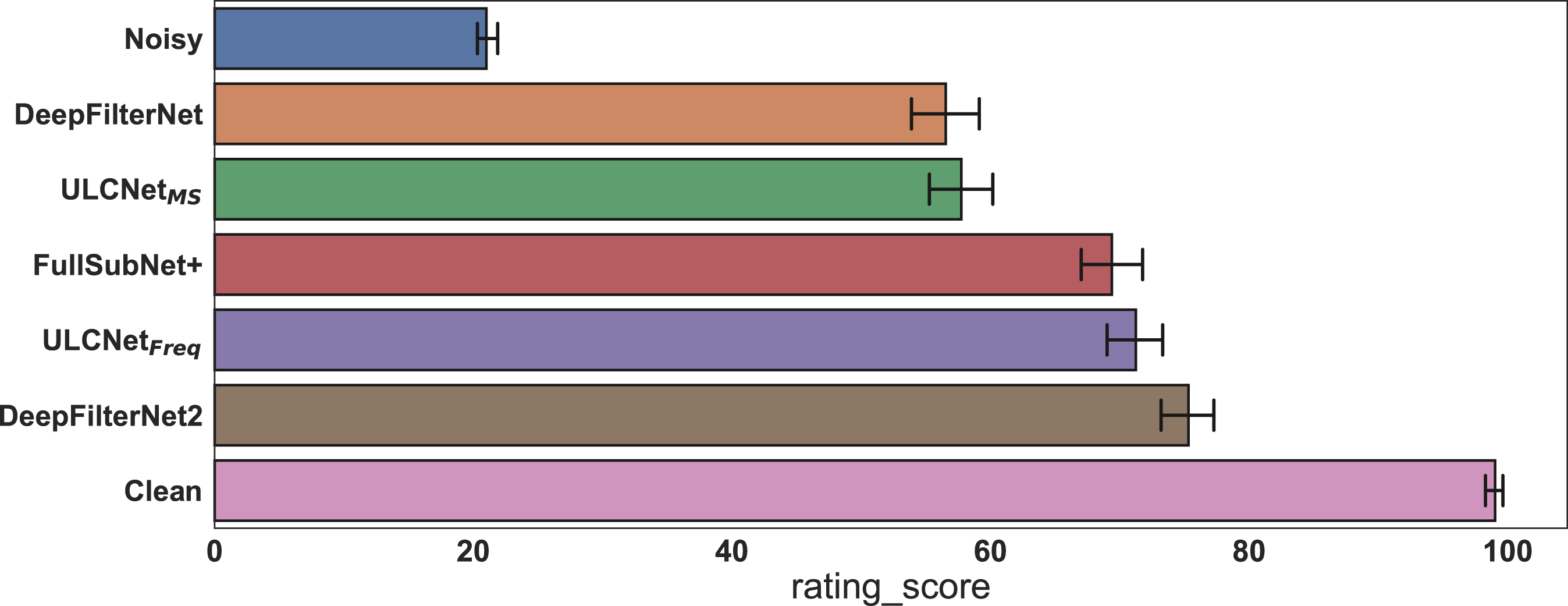, height=0.38\linewidth}
        \caption{\footnotesize MUSHRA Listening Test}
        \label{fig:subfig1}
    \end{subfigure}%
    \begin{subfigure}{0.5\textwidth}

        \epsfig{file=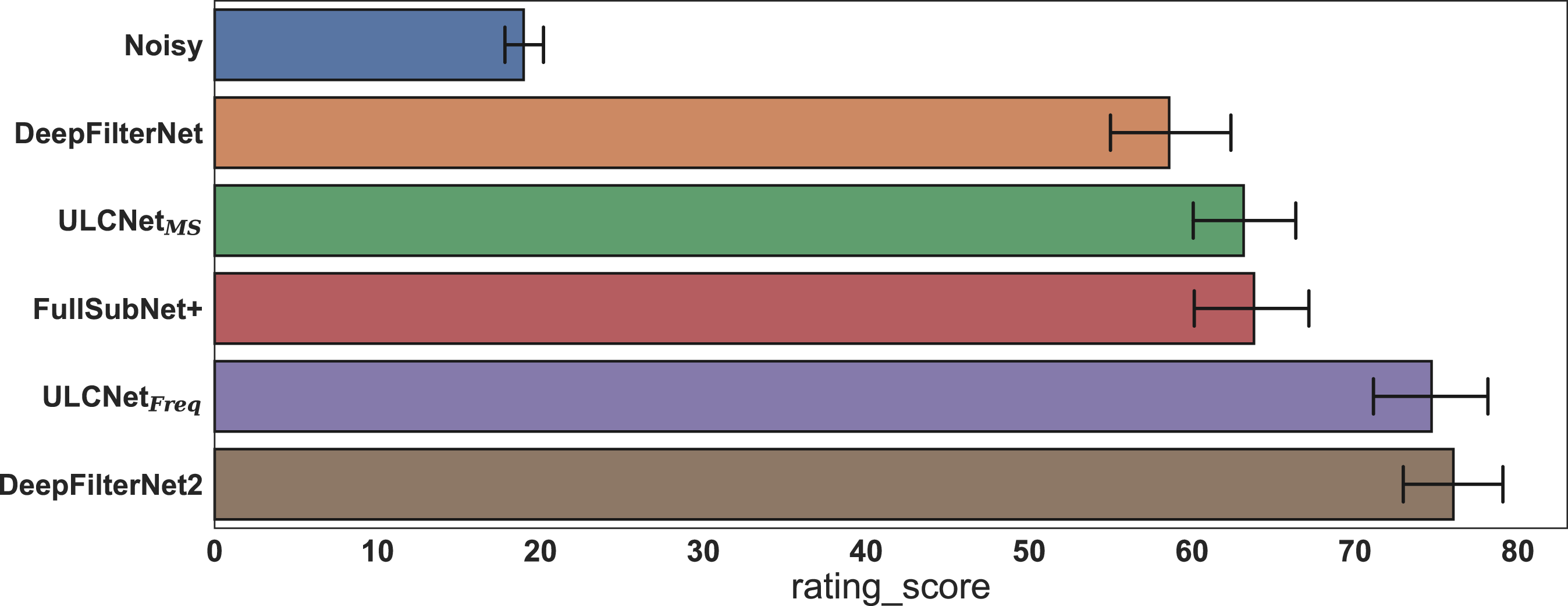, height=0.38\linewidth}
    
        \caption{\footnotesize Preference Test}
        \label{fig:subfig2}
    \end{subfigure}
    %\vspace{-.2cm}
    \vspace{-.6cm}\caption{\small Results for DNS challenge 2020 (a) non-reverb synthetic and (b) blind test set with errorbar representing the 95$\%$ confidence interval.}
    \label{fig:mainfig}
    \vspace{-.5cm}
\end{figure*}
\subsection{Implementation Details}
\label{sec:training}
\vspace{-0.1cm}
\subsubsection{Dataset}
\label{sec:data}
\vspace{-0.1cm}
In our experiments, we used the Interspeech 2020 DNS Challenge dataset \cite{reddy2020interspeech} for training. The
noisy mixtures in training and validation were generated at a 16 kHz sampling rate
by randomly selecting utterances from the clean speech set and the
noise set and mixing them at a random SNR between -10 dB and
30 dB. In total, we created a training dataset of around 1000 hrs. In $50\%$ of the training and validation dataset, we convolved clean speech
with a room impulse response (RIR) randomly selected from
the RIR set provided in \cite{reddy2020interspeech}. For data augmentation, we incorporated random low-pass filtering, upsampling, and different STFT windows.

\vspace{-0.3cm}
\subsubsection{Algorithmic Parameters}
\label{sec:pc}
\vspace{-0.2cm}
As outlined in section \ref{sec:methods}, our two-stage framework, operates in the TF domain after STFT transformation. For STFT, we adopt a window length of 32 ms, a 16 ms hop size, and FFT length of 512, yielding 257 frequency bins. For channel-wise feature reorientation, we apply an overlapping rectangular uniform window with a frequency resolution of 1.5 kHz (48 frequency bins) and an overlap factor of 0.33, leading to 8 subbands. The four separable convolutional layers in the Conv block, use a kernel size of $(1\times3 )$ to perform convolutions only along the frequency axis and use $\{ 32,64,96,128\}$ convolutional filters. Except for the first Conv layer, downsampling is achieved through max-pooling at a factor of 2 in the other three Conv layers. The FGRU layer comprises 64 units, succeeded by pointwise convolution with 64 filters. We use two subband temporal GRU blocks, each consisting of 2 GRU layers  with 128 units, which are then processed with 2 FC layers with 257 neurons. The second stage CNN architecture is composed of two 2D Conv layers with 32 filters and a kernel size of (1,3), followed by a pointwise Conv layer with 2 output channels which restores the desired output shape.
\vspace{-0.4cm}
\subsubsection{Loss Functions and Training}
\label{sec:Loss}
\vspace{-0.2cm}
We trained our proposed DNN model using various loss functions. Our experimental results indicated that the mean squared error (MSE) and multi-scale (MS) loss functions, as described in \cite{Choi2021RealTimeDA}, were more effective for this architecture. In this paper, we present the results of just two models trained with these two loss functions, each with a different power law factor. The first model denoted as ULCNet$_{\text{Freq}}$, utilizes a power law factor of $\alpha=0.3$ and is optimized with the MSE loss function. This loss is computed between the power law compressed clean speech signal $S$ and the estimated clean speech signal $\tilde{S}$. We do not combine the MS loss function with power law compression, as the cosine loss computation in the MS loss requires the power law decompressed time-domain clean and estimated speech signals, which would negate the effects of power law compression. The second model, referred to as ULCNet$_{\text{MS}}$, does not employ power law compression ($\alpha=1$). This model is trained exclusively with the MS loss function, as outlined in \cite{Choi2021RealTimeDA}. During training, we used the Adam optimizer with an initial learning rate (LR) of $4 \times 10^{-4}$ and a scheduler that reduces the LR by a factor of 10 every 3 epochs.

\vspace{-0.4cm}

\subsection{Results}
\vspace{-0.1cm}
\label{sec:results}
In our work, we evaluate our two proposed methods against five existing approaches from literature: NSNet2 \cite{braun2021towards,braun2020data}, PercepNet \cite{Valin2020APA}, FullSubNet+ \cite{chen2022fullsubnet+}, DeepFilterNet \cite{schroter2022deepfilternet} and DeepFilterNet2 \cite{schroter2022deepfilternet2}. The samples for objective and subjective evaluation were processed with code repositories mentioned on \cite{schroter2022deepfilternet2}. RTF was measured on a Core i5-8250U
CPU \cite{schroter2022deepfilternet2}.
\vspace{-0.3cm}
\subsubsection{Computational Complexity and Parameters}
\vspace{-0.2cm}
Our ULCNet models exhibit superior computational efficiency, as indicated by the MACS operations and RTF values presented in Table \ref{tab:Objectiveresult}. Notably, our proposed approach achieves significantly lower complexity and RTF, with just a quarter of the MACS operations compared to prior methods in the literature. In terms of model parameters, the ULCNet models are much smaller with 688K parameters, compared to the next best models, DeepFilterNet and DeepFilterNet2, which have 1.78M and 2.31M parameters, respectively.
\cite{schroter2022deepfilternet2}.

\vspace{-0.3cm}
\subsubsection{Objective Evaluation}
\label{sec:Objective Evaluation}
\vspace{-0.2cm}

For objective evaluation, we used PESQ \cite{union2007wideband} and SI-SDR \cite{le2019sdr} objective metrics. On the Voicebank+Demand test set, as shown in Table \ref{tab:Objectiveresult}, FullSubNet+ exhibits superior performance in SI-SDR metrics, surpassing all other methods. Our proposed ULCNet${_\text{Freq}}$ secures the second spot in SI-SDR improvement, although it suffers in PESQ metrics. Conversely, DeepFilterNet2 achieves the best PESQ scores but lags behind in SI-SDR metrics. ULCNet${_\text{MS}}$ achieves comparable results to other algorithms in both PESQ and SI-SDR metrics.

On the DNS challenge test set, FullSubNet+ outperforms all the algorithms in both PESQ and SI-SDR improvement. The remaining algorithms yield comparable results across SI-SDR and PESQ metrics, barring ULCNet$_{\text{Freq}}$, which again performs poorly in PESQ metrics.
\vspace{-0.4cm}
\subsubsection{Listening Test}
\label{sec:Listening Test}
\vspace{-0.2cm}
Objective metrics offer an initial comparative analysis of the
performance of different algorithms. However, recent studies have highlighted their limited correlation with human subjective ratings in noise suppression tasks, as individuals perceive noise and distortion uniquely \cite{reddy2021dnsmos,9914802}. Our own informal listening experience
aligns with this notion, as we find ULCNet$_{\text{Freq}}$ effective
in noise reduction despite its poor PESQ improvement.
Therefore, we conducted two listening tests to evaluate our proposed approaches compared to the best-performing
algorithms from Table \ref{tab:Objectiveresult}. In the listening test, we asked the listeners to rate the processing of different algorithms relative to each other considering the amount of noise reduction vs. speech distortion. In total 32 listeners took part in our listening test. The listener's group consisted of both expert and non-expert listeners.

%signal lowpass filtered with a cut-off frequency of 3.5 kHz. as PESQ is very sensitive to some of the distortions in speech \cite{torcoli2016effect}, which human listeners doesn't find so critical.

MUSHRA: In MUSHRA \cite{schoeffler2015towards} listening test, we selected 10 noisy samples from the DNS challenge 2020 test set representing various noisy scenarios. We used the unprocessed clean speech signal as the reference and the original unprocessed noisy signal as an anchor. The listeners were asked to rate the clean signal to 100 (max of MUSHRA scale). In the results as shown in Fig. \ref{fig:subfig1}, we find, DeepFilterNet2 and the proposed ULCNet$_{\text{Freq}}$ perform the best with mean rating scores of 75.46 and 71.72, respectively. FullSubNet+ performs better than the ULCNet$_{\text{MS}}$ and DeepFilterNet.

Preference Test: In the preference test, we selected 12 noisy samples from the DNS challenge 2020 blind test set. The aim of this test was to judge the performance of different methods blindly relative to each other. As the samples have no clean reference, we used the noisy sample as the hidden reference. Similar to MUSHRA results, here also DeepFilterNet2 and the proposed ULCNet$_{\text{Freq}}$ perform the best, as depicted in Fig. \ref{fig:subfig2}.  Proposed ULCNet$_{\text{MS}}$ achieves comparable results to FullSubNet+ and outperforms DeepFilterNet.
\vspace{-0.4cm}
\subsubsection{Discussion}
\label{sec:Discussion}
\vspace{-0.2cm}
In the listening test, we observed, that our newly proposed power law compression method helps in achieving better subjective perceptual quality in noise reduction. On the DNS challenge dataset, as illustrated in Fig. \ref{fig:subfig2} and Table \ref{tab:Objectiveresult}, our ULCNet${_\text{Freq}}$ model achieves notably higher subjective ratings (in contrast to the PESQ scores) compared to ULCNet${_\text{MS}}$, which employs no power law compression. This improvement is a result of our approach's more aggressive noise reduction strategy, which introduces some distortions that human listeners do not consider highly critical, as also evident from BAKMOS and SIGMOS metrics in Table \ref{tab:Objectiveresult}.

We also observed, that when the listeners had access to the original unprocessed clean signal, they became more discerning about residual noise and speech distortions. As a consequence, the gap in the mean subjective rating of ULCNet$_{\text{Freq}}$  and DeepFilterNet2 between the blind preference test and reference-based MUSHRA listening test, slightly increases from 1.34 to 4.07, respectively, as shown in Fig. \ref{fig:mainfig}. However ULCNet$_{\text{Freq}}$ performs overall quite satisfactorily and is comparable to DeepFilternet2 and clearly surpasses all other algorithms. For our listening test samples, please refer to \url{https://fhgainr.github.io/ULCNet/}.

To show the suitability of our ULCNet models for embedded devices, we also measured the RTF on a single core of a Cortex-A53 1.43 GHz processor, where it achieved 0.127 RTF (182MHz).

%Due to the low power consumption and computational efficiency of our approach, the proposed ULCNet$_{\text{Freq}}$ model is already deployed in two consumer-embedded devices.

\vspace{-0.4cm}
\section{Conclusion}
\vspace{-0.2cm}
In this paper, we demonstrate the feasibility of designing an ultra-low complexity DNN model for noise suppression with minimal performance degradation. Our proposed ULCNet$_{\text{Freq}}$ achieves subjective results on par with the SOTA DeepFilterNet2, requiring only $25\%$ of the computational complexity and model size. The modified power law compression method also shows benefits in enhancing the subjective perceptual quality. 

\blfootnote{This work has been supported by the Free State of Bavaria in the DSAI project.}

\vfill\pagebreak

\footnotesize
\bibliographystyle{IEEEbib}
\bibliography{strings,refs}

\end{document}